# A TEMPERATURE ANALYSIS OF HIGH-POWER ALGAN/GAN HEMTS

*Jo Das[1], Herman Oprins[1], Hangfeng Ji[2], Andrei Sarua[2], Wouter Ruythooren[1], Joff Derluyn[1], Martin Kuball[2], Marianne Germain[1] and Gustaaf Borghs[1]*

[1]Interuniversity Micro Electronics Center (IMEC), Kapeldreef 75, B-3001 Leuven, Belgium.
[2]H.H. Wills Physics Laboratory, University of Bristol, BS8 1TL, United Kingdom.

## ABSTRACT

Galliumnitride (GaN) has become a strategic superior material for space, defense and civil applications, primarily for power amplification at RF and mm-wave frequencies. For AlGaN/GaN high electron mobility transistors (HEMT), an outstanding performance combined together with low cost and high flexibility can be obtained using a System-in-a-Package (SIP) approach. Since thermal management is extremely important for these high power applications, a hybrid integration of the HEMT onto an AlN carrier substrate is proposed. In this study we investigate the temperature performance for AlGaN/GaN HEMTs integrated onto AlN using flip-chip mounting. Therefore, we use thermal simulations in combination with experimental results using micro-Raman spectroscopy and electrical dc-analysis.

## 1. INTRODUCTION

The AlGaN/GaN material system has many advantages for high power RF applications [1]. It combines a higher output power density (even at high frequency), a higher output impedance (easier matching), a larger bandwidth and better linearity than other existing technologies (like Si-LDMOS or GaAs p-HEMT). These features follow directly from the physical properties of GaN: the large bandgap ($E_g$ = 3.4 eV) results in a breakdown electrical field ten times larger than Si. This allows transistor operation at high bias voltage (>50V). Moreover it is possible to engineer piezoelectric AlGaN/GaN heterostructures, with spontaneous formation of a two-dimensional electron gas (2DEG), exhibiting high electron mobility and high current density, combined with high electron saturation velocity.

The AlGaN/GaN heterostructures are usually grown on sapphire, SiC and recently also on Si substrates. Sapphire substrates have the advantage of low cost, however, for realizing high power applications their low thermal conductivity (k = 35 W/m.K) compared to e.g. SiC (k = 400 W/m.K) is a major drawback. To overcome the low thermal conductivity of sapphire, flip-chip bumps can be used to interconnect the HEMT both electrically and thermally to a AlN carrier [2]. By integrating passive components onto the AlN, using a Multi-Chip-Module technology a System-in-a-Package (SIP) integration is obtained. In this paper we will study a new flip-chip bump design, to achieve a good thermal performance. We will compare finite element thermal simulations to experimental results, using electrical DC-characterization, and micro-Raman spectroscopy. We will demonstrate that using the novel bump design an improved thermal performance for multi-finger HEMTs on sapphire substrates can be achieved.

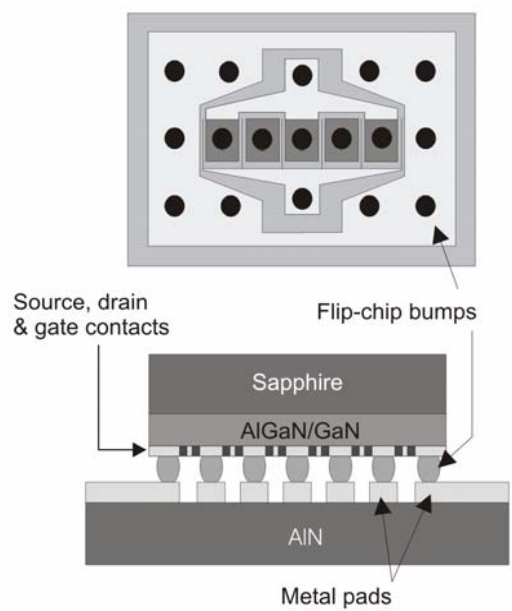

Fig.1: Top view and cross-section of the flip-chip bump design: bumps are placed directly on the source and drain ohmic contacts. In this way, there is a direct thermal and electrical contact between the ohmic contacts and the metal pads on the AlN carrier.

## 2. FABRICATION

The AlGaN/GaN heterostructures used in this work were grown by metal-organic vapor phase epitaxy (MOVPE) on sapphire substrates. On top of the AlGaN, an in-situ 3.5 nm $Si_3N_4$ passivation layer is grown. This $Si_3N_4$ layer enhances the electrical performance of the device, reduces the dispersion effects, and protects the surface during processing [3]. The HEMTs are fabricated using a 5-step process: mesa etching, ohmic contact formation, gate and contact metal deposition and passivation. The gate length $L_g$ is 1.5 μm, and the width of a single gate finger $W_g$ = 100 μm. The distance between two neighboring gate fingers is 100 μm.

To lower the device temperature, we integrated the HEMT onto a AlN carrier substrate using flip-chip bumps. In our new flip-chip bump design, bumps are placed both on source and drain contacts of the HEMT (see Fig.1). More details about the flip-chip bump design can be found in [4]. On the AlN substrate, metal contact pads are deposited by electroplating. Note that AlN is here the preferred carrier substrate, because of its high thermal conductivity and low substrate loss at RF frequencies.

TABLE I
THERMAL CONDUCTIVITIES USED IN THE SIMULATIONS

| Material | Thermal conductivity [W/m.K] |
|---|---|
| GaN | $160.(300\ K/T)^{1.4}$ |
| Sapphire | $35.(300\ K/T)$ |
| Au | 310 |
| AlN | $180.(300\ K/T)^{1.4}$ |
| Solder bumps | 50 |

## 3. THERMAL SIMULATIONS

To study the thermal behavior of the HEMT device for the different integration structures, a fully three dimensional finite element model is used [5]. Heat sources are defined between the ohmic contact regions of the multiple gate-finger HEMT. The heat generation is assumed to be uniform within these heat sources. For the simulations a constant temperature at the backside of the AlN carrier ($T$ = 25 °C) is defined as a boundary condition. The thermal conductivities of the materials used in the simulations are given in Table 1. For GaN, sapphire, and AlN, the temperature dependency of the thermal conductivity is taken into account with a temperature dependence as given in Table I.
Fig. 2 displays a cross-section of the temperature distribution within the HEMT after flip-chip integration,

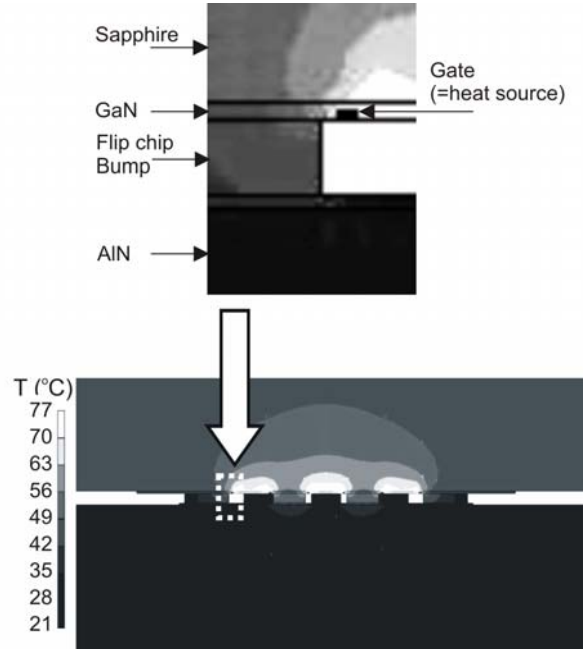

Fig.2: Temperature distribution simulation (cross-section) of a HEMT on sapphire after flip-chip, at a power dissipation of 3 W/mm.

for a total power dissipation of 1.8 W. Since the total gate width of the HEMT $W_g$ = 0.6 mm, this corresponds to a dissipated power density of 3 W/mm. From the simulations, we find a maximum temperature $T_{max}$ = 77 ºC, which is much lower compared to a HEMT on sapphire before integration ($T_{max}$ = 169ºC). From the figure it is very clear that the flip-chip bumps act as effective heat sinks. As a comparison, also a HEMT on SiC was simulated: since SiC has a much higher thermal conductivity, a lower maximum temperature was obtained: $T_{max}$ = 46 ºC. Nevertheless, this result shows that the thermal performance of the HEMT on sapphire after flip-chip becomes quite close to the HEMT on SiC, and at the same time it has the advantage of a much lower substrate cost.

## 4. EXPERIMENTAL RESULTS

DC-measurements and micro-Raman spectroscopy were used to assess the impact of flip-chip integration on the thermal performance of the HEMT. When a high drain-source voltage $V_{ds}$ is applied, the 2-DEG saturation current $I_{ds,sat}$ will decrease due to the self-heating effect. From this decrease ($\Delta I_{sat}$), it is possible to extract the temperature of the active region [6]:

$$\Delta I_{sat} = -g_m \cdot (I_{sat} \cdot \Delta R_s + \Delta V_T) \qquad (1)$$

In this equation, $\Delta R_s$ and $\Delta V_T$ represent the change of source resistance and the change of threshold voltage compared to their reference value at room temperature, respectively. The dependencies $V_T = f(T)$ and $R_s = f(T)$ were determined experimentally, using an external heating source to heat the devices up to constant temperature. As a result, the channel temperature versus dissipated power is obtained (see Fig. 3). From this figure it is clear that the flip-chip integration improves the thermal performance of the HEMT. Moreover there is a good agreement between the measured and simulated temperatures.

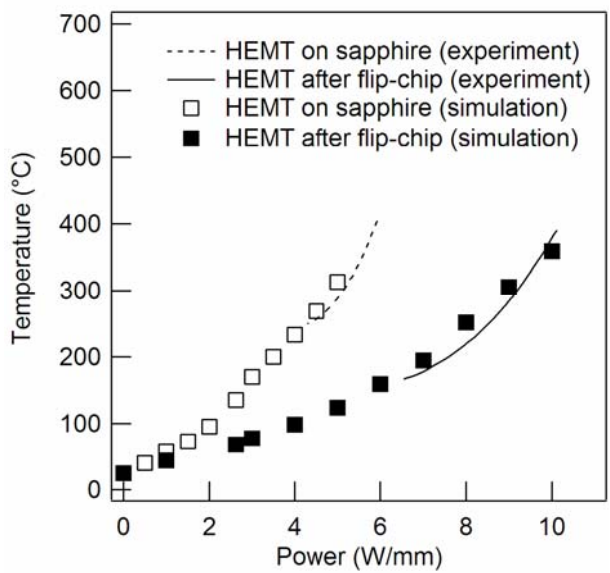

Fig.3: Temperature versus power for a HEMT on sapphire before and after flip-chip, obtained both from simulations and electrical measurements.

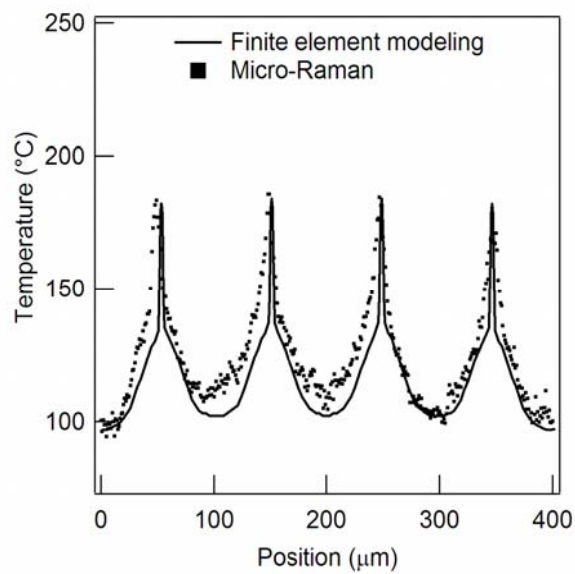

Fig.4: Temperature linescan of a HEMT on sapphire after flip-chip integration, at a dissipated power of 6.8 W/mm. Micro-Raman spectroscopy and finite element simulations are compared.

In order to measure experimentally the temperature distribution within the HEMT, an integrated Raman-IR system was used. This system consists of a Renishaw InVia Raman microscope equipped with motorized XY positioning stage and a QFI IR thermal microscope, switchable between Raman and IR detection modes. The 488nm line of an Argon ion laser was employed as excitation source for the Raman measurements. Raman shift of the $E_2$ phonon of GaN was probed to determine the temperature of powered devices. More details of the experimental set-up can be found in [7-8]. As the sapphire substrate was double-side polished, device temperature could be measured through the substrate. As a result, device temperature was accessible not only in the source drain openings, but also directly underneath the contacts and where the flip-chip tin bumps are located, which is a major advantage [9]. Fig. 4 displays an obtained linescan together with the thermal simulation showing that there is an excellent agreement between both obtained results. Note that peak temperatures determined from electrical device analysis (Fig. 3) are slightly smaller compared to peak temperatures obtained from the Raman measurements, possibly indicating that the electrical device characterization technique determines not peak device temperature, but a slightly lower average temperature.

## 5. CONCLUSION

We have investigated the temperature distribution in high-power AlGaN/GaN HEMTs before and after flip-chip, with thermal simulations, electrical and micro-Raman analysis. We obtained an excellent agreement between simulation and experimental results. We have shown that by using our new flip-chip design, with bumps on the source and drain ohmic contacts, a much better thermal performance is obtained, because the flip-chip bumps, which are very close to the active device area, act as very effective heat sinks.

## 6. ACKNOWLEDGEMENTS

The research at IMEC was supported by the ESA-ATHENA project under Contract No. 14205/00/NL/PA. The research at the University of Bristol was supported by EPSRC. H. Oprins thanks IWT for providing a scholarship. H. Ji thanks Universities UK for providing an ORS scholarship.

# 7. REFERENCES

[1] U. K. Mishra, P. Parikh, and Y.-F. Wu, "AlGaN/GaN HEMTs - an overview of device operation and applications", *Proc. IEEE*, vol. 90, pp. 1022-1031, Jun. 2002.

[2] J.J. Xu, S. Keller, G. Parish, S. Heikman, U.K. Mishra, and R.A. York, "A 3-10-GHz GaN-Based Flip-Chip Integrated Broad-Band Power Amplifier", *IEEE Trans. Microwave Theory Tech.*, vol. 48, pp. 2573-2578, Dec. 2000.

[3] J. Derluyn, S. Boeykens, K. Cheng, R. Vandersmissen, J. Das, W. Ruythooren, S. Degroote, M. R. Leys, M. Germain, and G. Borghs, "Improvement of AlGaN/GaN high electron mobility transistor structures by in situ deposition of a $Si_3N_4$ surface layer", *J. Appl. Phys.*, vol. 98, 054501, Sep. 2005.

[4] J. Das, H. Oprins, H. Ji, A Sarua, W. Ruythooren, J. Derluyn, M. Kuball, M. Germain and G. Borghs, "Improved Thermal Performance of AlGaN/GaN HEMTs by an Optimized Flip-Chip Design", accepted for IEEE Trans. Electron Devices.

[5] H. Oprins, J. Das, W. Ruythooren, R. Vandersmissen, B. Vandevelde, M. Germain, "Thermal analysis of AlGaN/GaN HEMTs", Proceedings International Workshop on Thermal Investigations of ICs and Systems, pp.71-75, (2005).

[6] J. Kuzmík, P. Javorka, A. Alam, M. Marso, M. Heuken, and P. Kordoš, "Determination of Channel Temperature in AlGaN/GaN HEMTs Grown on Sapphire and Silicon Substrates Using DC Characterization Method", IEEE Trans. Electron Devices, vol. 49, pp. 1496-1498, Aug. 2002.

[7] M. Kuball, A. Sarua, H. Ji, M.J. Uren, R.S. Balmer, and T. Martin, "Integrated Raman – IR Thermography on AlGaN/GaN Transistors", *2006 IEEE MTT-S International Microwave Symposium,* TH1B-06, Jun. 2006.

[8] M. Kuball, J.M. Hayes, M.J. Uren, T. Martin, J.C.H. Birbeck, R.S. Balmer, and B.T. Hughes, Measurement of temperature in high-power AlGaN/GaN HFETs using Raman scattering, *IEEE Electron Dev. Lett.,* vol. 23, pp. 7-9, Jan. 2002.

[9] H. Ji, A. Sarua, M. Kuball, J. Das, W. Ruythooren, M. Germain, and G. Borghs, "Flip-chip mounting for improved thermal management of AlGaN/GaN HFETs", *MRS proceedings*, vol. 892, 389-394 (2006).